\documentstyle[preprint,aps,psfig]{revtex} 
\begin{document}
\draft

\title{Ring current effects on the dielectric function of cylindrical nano-organic materials}

\author{St\'{e}phane Pleutin$^{1,2}$ and Alexander Ovchinnikov$^{2}$}
\address{$^1$ Fakult\"at f\"ur Physik, Albert-Ludwigs-Universit\"at-Freiburg, Hermann-Herder Stra\ss e 3, D-79104 Freiburg\\
$^2$Max-Planck-Institut f\"ur Physik Komplexer Systeme, N\"othnitzer
Stra\ss e 38, D-01187 Dresden}
\date{\today}
\maketitle
\begin{abstract}
We review recent results on the behaviour of the dielectric function of cylindrical nano-organic materials at very low frequencies in a magnetic field. For cylindrical structures - such as carbon nanotubes - the polarisability is shown to be a discontinuous function of a longitudinal magnetic field where plateau-like regions are separated by sudden jumps or peaks. A relation is pointed out between each discontinuity in the polarisability and the cross-over between ground and first excited states induced by the magnetic field. This one to one correspondence suggests to use measurements of the dielectric function in an applied magnetic field in order to obtain informations about the electronic structures of cylindrical nanostructures. In addition, it is shown, by studying finite graphene layers, that the measurement of the polarisability in a magnetic field could be a powerful way for detecting possible edge-states in amorphous carbon materials such as activated carbon fibres. Finally, the importance of the electron-electron interaction is emphasised by discussing examples of strongly interacting electrons on rings or cylinders, in the limit of infinite interaction.

\end{abstract}
\vskip2pc
\narrowtext
\section{Introduction}

It is known, since long time, that ring-shaped organic molecules, such as benzene or naphthalene, show very large diamagnetic susceptibilities and a very pronounced diamagnetic anisotropy: the diamagnetic susceptibility has normal values in the plane of the aromatic molecules but abnormally large ones in the direction perpendicular to this plane. A first qualitative explanation of this phenomenon as arising from the Larmor precession of electrons in orbits including many nuclei can be traced back to P. Ehrenfest in 1925 \cite{ehrenfest}. This idea was further developed eleven years later by L. Pauling who proposed a quasi-classical description of the $\pi$ electronic currents around the rings using an electrical network as a model \cite{pauling}. One year latter, in 1937, F. London set up the ``modern'' theoretical description of this
phenomenon by using a tight binding model which included the magnetic field in form of a complex phase of the usual transfer integral \cite{london}. This is the so-called London (or Peierls) substitution widely used nowadays. 

The comparison between the results given by these early theoretical works and the experimental data available at that time is quite satisfactory \cite{pauling,london}. However, the interest for this topic has nearly vanished most probably due to experimental limitations. For instance, in the theory of London, the electronic spectrum of aromatic molecules is a periodic function of the magnetic flux with a period equal to a flux quantum $\phi_0=hc/e$. But the value of the magnetic field needed to cover a full period is of order $10^5$ T for the materials existing until few years ago. Such huge values cannot be reached experimentally, and are due to the extremely small perimeter, of order a few tens of Angstr\"oms, of the materials studied previously.

Today, material science has made considerable progress and many new carbon materials with nanoscopic characteristic length exist. An example is given by multi-walled carbon nanotubes with a diameter of approximately 10 nm \cite{dresselhaus}. For these new materials, the magnetic field needed to reach a flux quantum is about a few tens of Tesla only and the periodicity mentioned before can be observed \cite{bachtold}.

On the other hand, during the last years considerable progress has been made in precise measurements of the real part of the low-frequency dielectric function $\epsilon^{\prime}(\omega)$. In particular these experiments have been extended to ultra-low temperatures. Ratios of $\delta \epsilon^{\prime}/\epsilon^{\prime}$ up to $10^{-7}$ were achieved in the mK temperature regime. A rather spectacular success associated with that progress was the observation of a strong magnetic-field dependence of the polarisability of multicomponent glasses \cite{glass}. These developments suggest reconsideration of the magnetic field effects in organic nanostructures such as large aromatic molecules or carbon nanotubes for instance. In particular, the magnetic field should have a very pronounced effect on the polarisability of these organic systems which should be observable nowadays \cite{fulde,pleutin1,pleutin2}. These measurements could then be used as a new spectroscopic tool to obtain informations on the electronic properties of organic materials. The aim of this paper is to review various theoretical works developed in Dresden during the last two years in that direction \cite{fulde,pleutin1,pleutin2,ahn}.

Similar ideas were already pointed out recently for quite different materials and purposes. Indeed, in \cite{efetov,bouchiat,blanter} small particles (rings, disks or spheres) were considered in connection with weak localisation. In particular, it was found that the polarisability should be greater in an applied magnetic field than in zero field because a field partially suppresses weak localisation. This theoretical prediction was observed very recently in the AC polarisability of mesoscopic rings \cite{deblock}.

Although the studies already done until now on the behaviours of the dielectric susceptibility of cylindrical organic materials under magnetic field \cite{fulde,pleutin1,pleutin2} should be improved in several ways by including, e.g., explicitely Coulomb interactions, we believe the phenomena pointed out here are robust in the sense that they will survive a more sophisticated theoretical treatments. Therefore, we think that the polarisability in a magnetic field could be a useful quantity to study experimentally in order to obtain insight into the electronic structure of carbon made materials. In our opinion, this justify the topic of this review and, we hope that it will help to motivate new experiments.

The series of works described here is related to the studies of persistent currents in metallic rings. Before starting, we would like to mention recent review articles on this topic. Interested readers could refer to Ref. \cite{eckern} for the case of diffusive electron motion and to Ref. \cite{richter} for ballistic electron motion.

The review is organised as follows. In section 2 we describe the Aharonov-Bohm effect in cylindrical nanostructures by using London's substitution. In section 3 and 4, we point out dramatic consequences of this effect on the dielectric function of nanoscopic organic cylinders such as carbon nanotubes (section 3), and a finite plane of graphite where the edge states existing in these materials define effective rings (section 4). Finally, before concluding, we point out a few properties of strongly interacting electrons on rings and cylinders in an applied field.

\section{Aharonov-Bohm effect in cylindrical nanostructures}
In this paper, we are interested in the effects of an applied magnetic field on the orbital degrees of freedom only. Therefore we consider spinless fermions except in the last section. The Zeeman interaction  was briefly discussed in \cite{fulde,pleutin1}. In the range of field strength we are interested in, the diamagnetic response is the most important one.

Let us first consider a ring of $N$ sites pierced by an axial magnetic
field. The Hamiltonian, making use of the London substitution \cite{london}, is the following

\begin{equation}
\label{Hring}
\hat{H}_R=t\sum^{N}_{n=1}(a^{\dagger}_{n+1}a_{n} e^{i\frac{2\pi}{N}\phi}+h.c.)
\end{equation} where $a^{\dagger}_{n}$, ($a_{n}$) are spinless fermion
creation (annihilation) operators and $\phi$ is the magnetic flux through
the ring in units of the flux quantum $\phi_0$ ($\phi_0=hc/e$). It is proportional to the area enclosed by the ring
\begin{equation}
\label{flux}
\phi=\frac{N^2a^2hc}{8\pi^2e}H
\end{equation} where $a$ is the lattice constant. For a ring with a few hundred sites and assuming $a\simeq 1.4 \AA$, the magnetic field needed to reach $\phi=1$ is not more than a few tens tesla - values reachable experimentally. 

Introducing the wave function
$|\Psi>=\sum_{n=1}^{N}c_n |\psi_n>$, where $|\psi_n>$ are the $\pi$ orbitals
located at sites $n$, one obtains the following eigenvalue equations with
periodic boundary conditions
\begin{equation}
\label{eigenring}
\left\{\begin{array}{l}
c_n \epsilon =
tc_{n+1}e^{i\frac{2\pi}{N}\phi}+tc_{n-1}e^{-i\frac{2\pi}{N}\phi}\\
c_{N+1}=c_1
\end{array}\right.
\end{equation}where $\epsilon$ is the eigenvalue. By performing a pseudo-gauge
transformation introduced by Bayers and Yang \cite{yang},
$\tilde{c}_n=c_n e^{i\frac{2\pi}{N}\phi n}$, Eqs. (\ref{eigenring}) can be written in the form
\begin{equation}
\label{eigenringGauge}
\left\{\begin{array}{l}
\tilde{c}_n \epsilon =
t\tilde{c}_{n+1}+t\tilde{c}_{n-1}\\
c_{N+1}=c_1e^{i2\pi\phi}
\end{array}\right.
\end{equation} The effect of the magnetic field
appears in the new periodic boundary conditions. An analysis of
them shows that the spectrum of the system is a periodic function of $\phi$ with period 1, as already mentioned in the introduction. This is an important information since it
implies that every thermodynamic quantity is also periodic function of
$\phi$ with the same period. Moreover, the system is invariant with respect to the transformation $H \rightarrow -H$. Therefore its spectrum - and every thermodynamic quantity - is symmetric with respect to $\phi=0$. In principle, it is then necessary to study the system only for $\phi \in [0,1/2]$.

The eigenvalues for the ring are
\begin{equation}
\label{eigenvaluering}
\epsilon_p=2t\cos(\frac{2\pi}{N}(p+\phi))
\end{equation}with $p=0,\pm1,\pm2,...$. Using this expression one notices crossing between levels occurring at special values of $\phi$, namely $\phi=0,1/2$ and $1$ (for an example see Fig. 1). 

Let us now consider cylinders made by rolled square lattices with an
applied axial magnetic field (Fig. 2). The Hamiltonian for spinless fermions is
\begin{equation}
\label{Hcylinder}
\hat{H_0}=t\sum^{N}_{n=1}(a^{\dagger}_{n+1,m}a_{n,m} e^{i\frac{2\pi}{N}\phi}+h.c.)+t\sum^{M}_{m=1}(a^{\dagger}_{n,m+1}a_{n,m}+h.c.)
\end{equation}where the magnetic flux is still given by expression (\ref{flux}). The resulting eigenvalues are

\begin{equation}
\label{spectrum}
\epsilon_{p,q}=2t\cos(\frac{2\pi}{N}(p+\phi))+2t\cos(\frac{\pi}{M+1}q)
\end{equation} with $-N/2 \leq p \leq N/2-1$ and $ 1 \leq q \leq M$. We have applied open
boundary conditions at the ends of the cylinder. The resulting one-electron wave functions are
\begin{equation}
\label{wavefunction}
|\Psi_{p,q}> = \sqrt{\frac{2}{N(M+1)}}
\sum_{n=1}^{N}\sum_{m=1}^{M}e^{i\frac{2\pi}{N}pn}\sin(\frac{\pi}{M+1}qm
)a^{\dagger}_{n,m}|0>
\end{equation} where $|0>$ is the vacuum.

Again, using the expression (\ref{spectrum}), one notices many level crossings induced by the magnetic field (Fig. 3). Compared to the results obtained for rings, the important difference is that the crossings appear for values of $\phi$ depending on the geometry of the system $(N,M)$; in particular, they appear for other values than $0$, $1/2$ and $1$. Due to their intricate patterns, numerous changes of the ground state induced by the magnetic field do occur. As an example, we show in figure 4 as function of the magnetic flux the energies of the ground state and the first excited state of a cylinder with $N=100$, $M=100$ and with $N_e=1000$ electrons in it. The number of crossing points and the values of $\phi$ for which they occur depend crucially on the geometry of the system as we have already mentioned above but, also, on the electron filling. We will not attempt to explain the crossing point pattern here. Instead, as shown in the next section, we want to point out that the changes of the ground state associated with the crossing points, have strong influences on the static polarisability. Therefore this quantity could be used as a new spectroscopic tool to obtain informations on excited states or on the system geometry for instance.

Carbon nanotubes form a class of cylindrical systems of growing importance nowadays \cite{dresselhaus,saito}. The calculations above can be straightforwardly extended to those systems, in particular for achiral nanotubes, i.e. the armchair and zig-zag nanotubes  considered here \cite{saito}. For these particular systems with a uniform magnetic field along their axis, we find by using again a nearest-neighbours tight binding model, the following spectrum \cite{saito,fulde,pleutin1}

\begin{equation}
\label{spectrumnano}
\epsilon_{p,q}=\pm (1+u_p \pm (u_pv_q)^{1/2})^{1/2}
\end{equation}where for zig-zag nanotubes
\begin{equation}
\label{zig-zag}
\begin{array}{c}
u_p=2(1+\cos(\frac{2\pi}{N}(p+\phi))),\quad \mbox{with} \quad p=1,..,N\\
\\
v_q=2(1+\cos(\frac{\pi}{M+1}q)),\quad \mbox{with} \quad q=1,..,M
\end{array}
\end{equation}and for armchair nanotubes
\begin{equation}
\label{armchair}
\begin{array}{c}
u_p=2(1+\cos(\frac{\pi}{M+1}p)),\quad \mbox{with} \quad p=1,..,M\\
\\
v_q=2(1+\cos(\frac{2\pi}{N}(q+\phi))),\quad \mbox{with} \quad q=1,..,N
\end{array}
\end{equation}The roles of $u_p$ and $v_q$ are just exchanged from one case to the
other.

Using the expressions (\ref{spectrumnano}-\ref{armchair}), one notices that - as for the case of rolled square lattices - the ground state of rolled honeycomb lattices changes also strongly in a magnetic field. Therefore, as for the case of rolled-square lattices, the static electric polarisability of carbon nanotubes are also expected to be strongly influenced by an axial magnetic field \cite{fulde,pleutin1}. It is the purpose of the next section to analyse the structure of the dielectric susceptibility of cylindrical organic nanostructures in presence of an applied magnetic field. Moreover, the electronic spectrum of carbon nanotubes is also a periodic function of the magnetic flux, a behaviour which has been confirmed experimentally in \cite{bachtold}.

\section{Static electric polarisability of cylindrical nanostructures under magnetic field}

Very precise measurements of the polarisability can be done nowadays. Two recent examples are given in \cite{glass} for multicomponent glasses and in \cite{deblock} for mesoscopic metallic rings. The complicated electronic spectrum of cylindrical organic nanostructures - including carbon nanotubes - induced by an axial magnetic field causes important effects in the polarisability of these systems as we want to discuss in the following.

For simplicity let us focus on rolled square lattices only. We give one example for carbon nanotubes in this section but, more informations can be found in \cite{fulde,pleutin1}. 

With a uniform DC electric field $E$, the tight binding model becomes
\begin{equation}
\label{TBE}
\hat{H}=\hat{H}_0+E\hat{d}
\end{equation}where $\hat{d}$ is the dipole operator. In the following, two different orientations for the electric-field are considered. The corresponding expression for the dipole operator are
\begin{equation}
\label{dipoleoperatorL}
\hat{d}= ea\sum_{n,m} (n-\frac{M}{2})a^{\dagger}_{n,m}a_{n,m}
\end{equation} for longitudinal field, where $a$ is the lattice spacing and $L=Ma$ is the total length of the tube, and
\begin{equation}
\label{dipoleoperatorT}
\hat{d}= eR\sum_{n,m}\cos(\frac{2\pi}{N}n)a^{\dagger}_{n,m}a_{n,m}
\end{equation} for transversal field, where $R$ is the radius of the cylinder (Fig. 2). The first case was studied in \cite{fulde}, the second in \cite{pleutin1}.

We are interested in the behaviour of the induced dipole moment as function of the magnetic flux
\begin{equation}
\label{dipole}
P(T,\phi)=\frac{1}{E}\frac{Tr \hat{d}e^{-\beta(\hat{H}-\mu)}}{Tre^{-\beta(\hat{H}-\mu)}}
\end{equation}where, as usual, $\beta=\frac{1}{k_B T}$ and $\mu$ is the chemical potential which is a function of the magnetic field.

We summarise in the following, the main results of the studies done in \cite{fulde,pleutin1}.

\subsection{Longitudinal electric field: Semi-classical approximation}

In this case, the Hamiltonian of the system is given by Eqs. (\ref{TBE}) and (\ref{dipoleoperatorL}). With this model it becomes tedious to evaluate the polarisability (\ref{dipole}), for long chains. Instead of doing exact calculations we proceed by using a semi-classical approximation very powerful in this specific context \cite{fulde}.

We are interested in the linear regime where the potential induced by the electric field, $eaEn$, is slowly varying on the scale of the Fermi wave length. Then, the Thomas-Fermi approximation may be applied \cite{mermin}. We assume that it is meaningful to consider an energy depending on the position, $m$, on the chain
\begin{equation}
\tilde{\epsilon}_{p,q}(m)=\epsilon_{p,q}+eaE(m-\frac{M}{2})
\end{equation} where $\epsilon_{p,q}$ is given by the expression (\ref{spectrum}). Next, within this semi-classical approximation, the polarisability is evaluated as follows.
\begin{equation}
\label{TFCylinder}
P(T,\phi)=\frac{ea}{EM}\sum_{p,q,m}\frac{(m-\frac{M}{2})}{e^{\beta(\tilde{\epsilon}_{p,q}(m)-\mu)}+1}
\end{equation}

At this stage, it should be noticed that the interaction which leads to an energy splitting at crossing points when the electric field is turned on, is neglected here. Therefore, a possible first order Stark effect near those points is not taken into account with this approximation.

To illustrate the relevance of this approximation, we consider first a linear chain in the absence of a magnetic field. In this case, the quasi-classical energy at position $m$ is given by
\begin{equation}
\tilde{\epsilon}_q(m)=2t\cos (\frac{\pi}{M+1}q)+eaE(m-\frac{M}{2})
\end{equation} and the corresponding polarisability by
\begin{equation}
\label{TFChain}
P_{lin}(T)=\frac{ea}{EM}\sum_{q,m}\frac{(m-\frac{M}{2})}{e^{\beta(\tilde{\epsilon}_{q}(m)-\mu)}+1}
\end{equation}

Figure 5 shows an example for a linear chain with 201 sites. The exact polarisability is compared with the semi-classical one (Eq. \ref{TFChain}) as function of the electron filling for two different values of the temperature. One can see that the deviations caused by the semi-classical approximation are less than $1\%$ or $\frac{1}{M}$. This justifies the use of this approximation for cylindrical systems in a magnetic field.

Figure 6 shows the results for a cylinder with $N=200$ sites along the circumference and $M=1000$ sites along the cylindrical axis. The temperature is $k_bT=10^{-4}$ in units of $t$ (around 1K) and the density is 0.74 electrons per site. Since the polarisability is symmetric with respect to the point $\phi=1/2$, as mentioned before, only the regime $0<\phi<1/2$ is shown. Obviously there is a huge effect of the magnetic field on the polarisability.

These conclusions are based on a semi-classical approximation. In the next subsection we present full quantum calculations for the case where the electric field is perpendicular to the cylindrical axis. Therefore, we postpone a complete discussion of the results to the next subsection where they do not suffer from any approximation in the context of our simplified model. 

\subsection{Transversal electric field: Quantum calculation}

The Hamiltonian describing the system is now given by Eqs. (\ref{TBE}) and (\ref{dipoleoperatorT}). The advantage of this choice is to allow for a separation of variable: the motion of the electrons along the circumference and the cylindrical axis can be treated separately. Exact calculations can be done even for very long cylinders. Proceeding that way, the problem is reduced to a study of one-dimensional rings in applied electric and magnetic fields \cite{pleutin1}.

We show in Fig. 7b the magnetic field dependence of the polarisability for
a cylinder with $N=101$ and $M=100$ and very few electrons on it, $N_e=100$
(this corresponds to a band filling of only $1 \%$). In this example, the electric field is such that
$eaE=10^{-3}t$ (linear regime) and the temperature is $k_bT=10^{-5}t$ ($T\simeq 0.1K$). We choose this example
because it shows very clearly the main behaviours of the dielectric response. In particular we choose $N=101$ because this gives an
illustration of the signature of the linear Stark effect. Together with this example the energies of the ground state and first excited state are shown without electric field as function of the magnetic flux (Fig. 7a).

First of all the induced
dipole moment as function of the magnetic field is periodic with a period of
$1$ and is symmetric with respect to $1/2$; these symmetries are
already apparent in the spectrum (\ref{spectrum}). Second, the induced
dipole moment shows clearly two main characteristics: (i) we can
notice the presence of small peaks at $\phi=0, 1/2, 1$, (ii) the induced dipole moment is a discontinuous function showing
several jumps separating plateau-like sections. Note that there is a slight curvature in the whole
spectrum, which is related to the persistent current induced by the
magnetic field. 

As noticed before, a magnetic field induces crossing between the
energies of the ground state and the first excited state (figure 4 and 7a). As it
can be seen in figures 7, there is a one to
one correspondence between level crossing at zero electric field and each
kink in the induced dipole moment. 

At each crossing, the ground state of the system changes. These two states
which are crossing respond differently to an
applied electric field. Both produce a quadratic Stark effect but generally of
different size. This explains why the
induced dipole moment is not a continuous function of the magnetic field. 

Near the crossing points, the response of the system will depend on whether or
not there is a nonzero matrix element of the electric field between the two states
involved. If the matrix element vanishes, the picture described above is
valid. On the contrary, if there
is an interaction between those states, due to the degeneracy,
the response will result in a linear (instead of quadratic) Stark effect with peaks of the induced
dipole moment. More precisely, using the expression of the wave function
(\ref{wavefunction}), the matrix elements of the dipole operator are calculated as
\begin{equation}
\label{selectionrules}
< \Psi_{p,q}|\hat{d} | \Psi_{p',q'}>=eR \delta_{p',p\pm1} \delta_{q',q}
\end{equation} With this equation, it is easy to see that a linear Stark effect
can occur only for a  magnetic flux of $\phi=0,1/2,1$. These are precisely the values for which one obtains peaks in our first example (Fig. 7b).

It is worth noticing that large enough Coulomb interactions can
change drastically the selection rules (\ref{selectionrules}). The same is true for distortions of a  cylinder as it was shown in \cite{pleutin1} for elliptical cylinders. Appearance of
a linear Stark effect for different values of the magnetic flux could then give a way
to quantify importance of electronic correlation effects or to characterise the geometry of the systems under consideration.

At low enough temperature, one deals essentially with the spectroscopy of a few
levels around the Fermi level. Therefore it is not surprising that the induced
dipole moment for a system which is well ordered, well oriented and of well
defined size ($N,M$), depends strongly on the electronic
density. This is clearly seen from Figs. 8a and 8b. They are for the
very same system considered before but for three different electron fillings
$N_e=100, 101$ and $102$. At such a low density, the relevant mean level
 spacing, $\Delta E$, behaves like $\Delta E \approx 1/N^2$. Figure 8a is for a
temperature lower by one order of magnitude than $\Delta E$, while Fig. 8b is
for a temperature higher by one order of magnitude than $\Delta E$. With
a typical value for $t$ ($t\approx 2 eV$) one can estimate a temperature of
$10^{-1}K$ for the spectrum 8a and $10K$ for the spectrum 8b. With these
figures we want to emphasise the unique sensitivity of the proposed measurements
and also that it is the necessary to work at very small temperatures in order to
extract maximum informations. 

The last examples in Fig. 9 is an illustration showing that the same kind of behaviour is expected in carbon nanotubes. It is for armchair (Fig. 9a) and zig-zag (Fig. 9b) nanotubes, when both are $1 \%$ away from half-filling and at $T \simeq 0.1K$ \cite{pleutin1}.

Today it is possible to measure accurately very small variations in the real part of the dielectric function \cite{glass,deblock}. Therefore, the dramatic magnetic field effects on the static polarisability of mesoscopic cylindrical systems, discussed here, could be measured and analysed. Several important informations about excited states could then be
obtained. First, the positions of the discontinuities
and peaks should give informations about the energies of the excited states. The nature of the
response - linear or quadratic Stark effect could be detected and therefore
should give informations about the symmetry of the excited states. The
magnitude of the response should also give informations about the coupling
constants. Finally, the different curvatures observed in the whole spectrum
could give informations related to the persistent current - or ring current - induced by the
magnetic field. 

Until now we were discussing cylindrical systems, such as carbon nanotubes, with a diameter in the mesoscopic range. The diamagnetic current induced by the applied magnetic field is a ring current in a strict geometrical sense. In the following, we will discussed quite different systems closer to those discussed by Ehrenfest, Pauling and London \cite{ehrenfest,pauling,london}. Indeed, we consider finite graphene planes where the magnetic field induces a ring current running along the perimeter of these systems. It is very close to the earlier studies where ring currents in benzene or naphthalene molecules were considered. The mesoscopic graphene sheets are new materials of importance for which the magneto-polarisability \cite{bouchiat} could give very useful informations.  

\section{Static electric polarisability under magnetic field of nanographite layers}

Very recently a new type of magnetism similar to a spin glass state was found
for activated carbon fibres prepared at heating temperature over 1200\char23 C
\cite{acf}. These materials are supposed to be well described by disordered networks of nanographite layers with a characteristic length of some tens of Angst\"oms \cite{dresselhaus}. It is worth noting that further increase of the heating temperature will drive the system towards complete graphitisation. 

On the other hand, recent theoretical works on graphite ribbons have shown the existence of edge states whose wave functions are localised in the near vicinity of the edges of the ribbons and whose energies form a band with a small dispersion centred in the middle of the gap \cite{edge2,sigrist}. Therefore, these states can have important physical consequences since they make a relatively large contribution to the density of state at the Fermi level. For instance, they are expected to affect novel electrical and magnetic properties such as a paramagnetic behaviour at low temperature \cite{sigrist}. In the light of these theoretical works, it was suggested by Shibayama et al. that the edge states should be responsible for the spins observed in \cite{acf}. The spin glass behaviours could then be explained by invoking Heisenberg like interactions between these effective spins since the disordered network of nanographite layers gives rise to a random exchange interaction network.

In this context, we want to stress in this section that measurements of the magneto-polarisability may be a very direct and efficient way to detect possible edge-states in activated carbon fibres or any other amorphous carbon
material. This new way of probing electronic structure of carbon materials could help to support the hypothesis made in \cite{acf}. Moreover, we believe that this quantity could give a way to access characteristic length of graphitic regions in such compounds - quantity studied today using Raman spectroscopy \cite{amorphous}.

For that purpose, we consider first rectangular finite planes of
graphite (Fig. 10). These simplest systems show two different kinds of edges: zig-zag and armchair edges. The one-band $\pi$ electrons are again described by a nearest-neighbours tight binding Hamiltonian defined on the topologically equivalent brick-type lattice already used in \cite{sigrist},

\begin{equation}
\label{tbgraphene}
\hat{H}=\sum_{<(n,m),(n',m')>}ta^{\dagger}_{n,m}a_{n',m'}
\end{equation}where the summation is over nearest-neighbour sites located at $(x,y)$ coordinates, $(n,m)$ and $(n',m')$.

As found in references \cite{edge2,sigrist}, our studies of (\ref{tbgraphene}) show the existence of edge states for finite rectangles of graphite. Moreover, in agreement with the results of \cite{edge2,sigrist} for ribbons, the wave functions of the edge states are shown to be localised preferentially on the zig-zag edges. In particular the weight of these functions is negligible on the armchair edges. This remark could be of importance for real materials leading to the question about the stability of such states in layers with disordered shape. We postpone a discussion of this important question to a future publication. Finally, the energies of these states are near the middle of the gap; the closer it is to the middle, the more localised on the edges is the corresponding wave function.

Next we apply a uniform magnetic field $\vec{H}$ perpendicular to the layer. In the tight binding model we then proceed once more by using the London substitution \cite{london}, $t \rightarrow t_{(n,m)\rightarrow (n',m')}=te^{i2\pi \frac{e}{ch}\int^{(n',m')}_{(n,m)}\vec{dl}\vec{A}}$, where $\vec{A}$ is the vector potential. We write
\begin{equation}
\label{twithH}
\begin{array}{c}
t_{(n,m)\rightarrow (n+1,m)}=te^{i\gamma_x(n,m)}\\
t_{(n,m)\rightarrow (n,m+1)}=te^{i\gamma_y(n,m)}
\end{array}
\end{equation}where $\gamma_x(n,m)$ and $\gamma_y(n,m)$ are the appropriate line integral of the vector potential. Since the magnetic field is uniform, one must have (cf Fig. 10), again in units of $\phi_0$
\begin{equation}
\begin{array}{l}
\gamma_y(n,m)-\gamma_y(n+2,m+1)+\\
\quad \quad \gamma_x(n,m+1)+\gamma_x(n+1,m+1)-\gamma_x(n+1,m)-\gamma_x(n,m)=2\pi\phi.
\end{array}
\end{equation}

For calculations we
choose, as in \cite{sigrist}, the Landau gauge.
\begin{equation}
\begin{array}{c}
\gamma_y(n,m)=0\\
\gamma_x(n,m)=\pi\phi m
\end{array}
\end{equation}
 It should be noticed that the elementary magnetic flux is defined here with respect to the size of a unit cell, namely a benzene ring, which requires an extremely large field except for the edge states where the relevant unit is given by the size of the whole layer - corresponding to a much more reasonable magnetic field of some tens Tesla. In that sense, the edge-states define an effective rings. The idea is quite simple: only the edge states will react to small values of an applied magnetic field such that the corresponding characteristic length of the cyclotron radius is of the same order than the size of the layer. Subsequently, the signature of the edge states should be seen in the polarisability as function of magnetic field.

We then add a static uniform electric field in the plane of the layer and calculate the magneto-polarisability for different electron-fillings. In this case, the dipole operator takes the following general expression $\hat{d}=e\sum_i \vec{r}_i a_i^{\dagger}a_i$ where the site vector $\vec{r}_i$ is defined with respect to an arbitrary origin. We choose a system with a total of 76 benzene rings with 10 along the zig-zag edges. We work at small temperature $k_BT=10^{-5}t_{ij}$ which could be estimated to $T \simeq 0.1K$.

The particular example studied here shows only two nearly degenerate edge states at zero energy. The edge states  will be populated at half-filling but since the temperature is low compared to the energy gap, they are not occupied if one reduces the electron number by one. Figure (11) shows the magneto polarisability for these two cases. We see very clearly a quadratic dependence of the polarisability as function of the magnetic field for the half-filled case only. For the other case, no sensible effect are seen. We can then conclude that the edge states are responsible for this quadratic behaviour in accordance with the argument given above.

To conclude this section, it should be stressed that the relative variations of the polarisability in Fig. 11 for the case with occupied edge states can be detected nowadays. Therefore, we believe it is possible to identify the existence of edge states in activated carbon fibres, or other amorphous carbon materials, using measurements of the magneto-polarisability. Moreover, we believe that a careful analysis of the quadratic behaviour shown in Fig. 11 can help to gain important informations about these disordered materials such as the effective sizes of the graphite-like regions for instance.

\section{Ring-current in strongly correlated cylindrical nanostructures}

So far we didn't consider Coulomb interactions. Instead, we were assuming that they were included in the effective one-electron parameters of our tight-binding models. However, for quantitative results of the polarisation, it is needed to include explicitely those interactions. It is well known that the absolute value of the polarisation of organic compounds decreases strongly by including interaction corrections \cite{louie}. We will not attempt such studies here which will be the subject of a subsequent publication. Instead, we will consider several strongly interacting systems, rings or small cylinders, in the extreme limit of infinite on-site interactions, our purpose being to point out that Coulomb interactions can produce not only quantitative changes but, also, important qualitative changes in the behaviour of the polarisability.

\subsection{Fractional Aharonov-Bohm effect}

In the early nineties, several authors have studied the behaviour of persistent currents in mesoscopic Hubbard rings threaded by a magnetic field \cite{kusmartsev,fab}. One of the most prominent result of these studies was obtained in the limit of infinite interaction, $U\rightarrow \infty$. In this limit, Kusmartsev pointed out that, in the case of spin $1/2$ fermions, the persistent current has a surprising short period given by $\Delta \phi=\frac{2 \pi}{N_e}$ (in unit of $\phi_0$), where $N_e$ is the number of electrons \cite{kusmartsev}. He called this intriguing new phenomena, fractional Aharonov-Bohm effect. It is this result which we shortly describe here.

The tight binding Hubbard model for a ring with $N$ sites is of the form

\begin{equation}
\label{hubbard}
\hat{H}=t\sum_{n,\sigma}(a^{\dagger}_{n+1,\sigma}a_{n,\sigma}e^{i\frac{2\pi}{N}\phi}+h.c.)+U\sum_n \hat{n}_{n,\uparrow}\hat{n}_{n,\downarrow}
\end{equation} where $\hat{n}_{n,\sigma}$ is the usual electron number operator. The magnetic flux $\phi$ is given by the expression (\ref{flux}).

An advantage of the Hubbard ring is to be one of the strongly-correlated-electron system that is solvable by using the Bethe ansatz \cite{lieb}. In the presence of a magnetic field the form of the wave function doesn't change

\begin{equation}
\label{betheansatz}
\Psi(x_1,x_2,...,x_{N_e})=\sum_P[Q,P]\exp(i\sum_{j=1}^{N_e}k_{P_j}x_{Q_j})
\end{equation}where $P=(P_1, P_2,..., P_{N_e})$ and $Q=(Q_1, Q_2,..., Q_{N_e})$ are two permutations of $(1, 2,..., N_e)$. The coefficients $[Q,P]$ as well as $(k_1, k_2,..., k_{N_e})$ are determined from Bethe's equations which in a magnetic field are modified by the addition of the flux phase $\phi$ \cite{kusmartsev,fab}.
\begin{equation}
\label{betheequation1}
\exp(ik_jN-2\pi \phi)=\prod_{\beta=1}^{N_s}(\frac{it \sin k_j -i\lambda_{\beta}-U/4}{it \sin k_j -i\lambda_{\beta}+U/4})
\end{equation}
\begin{equation}
\label{betheequation2}
-\prod_{j=1}^{N_e}(\frac{it \sin k_j -i\lambda_{\beta}-U/4}{it \sin k_j -i\lambda_{\beta}+U/4})=\prod_{\alpha=1}^{N_s}(\frac{i\lambda_{\alpha}-i\lambda_{\beta}+U/2}{i\lambda_{\alpha}-i\lambda_{\beta}-U/2})
\end{equation}$N_e$ is the number of charges i.e. the number of electrons, and $N_s$ is the number of down-spins. The eigenstates of a ring are characterised by the momenta $k_j$ of the charges and the rapidities $\lambda_{\beta}$ of spin waves. The energy of the system is given by
\begin{equation}
\label{energybethe}
E=2t\sum_{j=1}^{N_e}\cos k_j
\end{equation} 

In the limit of infinite interaction $U\rightarrow \infty$ Bethe's equations are greatly simplified to give the following results

\begin{equation}
\label{uinfinity}
k_j=\frac{2\pi}{N}(I_j+\phi+\frac{1}{N_e}\sum_{\alpha=1}^{N_s}J_{\alpha})
\end{equation} where the quantum numbers $I_j$ and $J_{\alpha}$ are connected with charge and spin degrees of freedom respectively. They are either integers or half odd integers, depending on the parities of the numbers of down- and up-spin electrons:
\begin{equation}
\label{quantumnumbers}
I_j=\frac{N_s}{2}(mod 1), \quad J_{\alpha}=\frac{N_e-N_s+1}{2} (mod 1).
\end{equation}

The equation (\ref{uinfinity}) is identical to the one describing a set of non-interacting spinless fermions on a ring threaded by a flux $\phi+\frac{1}{N_e}\sum_{\alpha=1}^{N_s}J_{\alpha}$. The second term of this expression is the effective magnetic flux created by the spin excitations. It comes from the fact that there is an additional phase shift between the momenta of the different electrons due to the repulsive interaction. Because of this shift, the periodicity of the Aharonov-Bohm flux takes a fractional value at the infinite interaction limit. It is easy to check this statement by using Eqs. (\ref{energybethe}) and (\ref{uinfinity}) and by appropriately choosing the quantum numbers (\ref{quantumnumbers}) in order to minimise the energy of the system. One example is shown in Fig. 12a for a ring with 8 sites and 2 electrons, one up and one down. Another example is shown in Fig. 12b for the very same ring but for 5 electrons, 3 up and 2 down. The fractional Aharonov-Bohm effect is clearly seen in these examples.

It is important to stress once more that the spin plays an essential role in this phenomena. This can be seen in (\ref{uinfinity}) where the quantum numbers associated to the spin degree of freedom appear as responsible of the fractional AB effect. Without this term we recover the known result for free fermion rings (\ref{eigenvaluering}). In particular, for spinless fermions, the periodicity with the flux is 1. 

In the case of infinite interaction or spinless fermions, we can use the following argument to obtain the correct flux periodicity and which stresses the important role played by the spin. Let us take a particular configuration with $M$ electrons, $|x_1,x_2,...,x_M>$. One may ask the question of how many hops of the particles in the same direction should take place in order to recover the starting configuration. In mathematical terms we may write

\begin{equation}
\label{periodicity}
(e^{i\frac{2\pi}{N}\phi}\hat{T}_1)^{N_1}(e^{i\frac{2\pi}{N}\phi}\hat{T}_2)^{N_2}...(e^{i\frac{2\pi}{N}\phi}\hat{T}_M)^{N_M}|x_1,x_2,...,x_M>=\alpha |x_1,x_2,...,x_M>
\end{equation}where $\alpha$ is a phase term cause by the Pauli's principle. $x_i$ are general coordinates of the electron $i$ taking account of the spin when necessary. $\hat{T}_i$ are the operator of translation by one site in the clockwise direction (for instance) of the electron $i$. $N_i$ is the number of one-site-hopping of the particle $i$. The answer to this question depends on whether or not we consider the spin degree of freedom.

In the cases we are interested in, the fermions are locked together and form a kind of dynamical Wigner crystal. By moving on the lattice, an electron will push the others in the same direction. For the case of spinless fermions, since all particles are indistinguishable, it is easy to see that $N_1+N_2+...+N_M=N$ to get back to the initial configuration. The same is true for an assembly of free spin $1/2$ fermions. On the other hand, one come back to the same configuration by interchanging the particles which, due to the Pauli's principle, give the following statistical flux: $\alpha=-e^{i\pi M}$. The immediate conclusions are, first, that the periodicity of the energy with the flux will be 1 (as for the free electron case) and, second, there will be a parity effect which discriminates between the cases with even and odd number of particles. Indeed, it was shown by Kusmartsev \cite{kusmartsev} (and references therein), that a system of interacting spinless fermions on a ring behaves as a paramagnet for even number of particles and as a diamagnet for an odd number of particles. This property is sometimes known as {\it Leggett's conjecture} \cite{leggett}.

 For the case of spin $1/2$ fermions, the conclusions are completely different. Indeed, one easily sees that $N_1+N_2+...+N_M=N.M$ and $\alpha=1$. Each particle should go round the ring and therefore, the statistical flux is just unity. Consequently, the effective charge of the system will be $Me$ and the periodicity of the energy will be $1/M$ - this is the fractional Aharonov-Bohm effect discovered by Kusmartsev \cite{kusmartsev}. Moreover, there is no parity effect: indeed, it is known that the system is always diamagnetic for infinite interactions \cite{kusmartsev2}.

On the basis of these analysis, let us now consider strongly-interacting cylindrical systems in some limiting cases where conclusions could be reach following the same argumentation.

\subsection{Strongly interacting rings with one electron per ring}
Let us first consider a stack of isolated rings with only one electron per ring. The rings are interacting following the model
\begin{equation}
\label{hubbardc}
\hat{H}=t\sum_{n,m}(a^{\dagger}_{n+1,m}a_{n,m}e^{i\frac{2\pi}{N}\phi}+h.c.)+U\sum_{n,m} \hat{n}_{n,m}\hat{n}_{n,m+1}
\end{equation} where there is a nearest-neighbour interaction between two electrons when they are on top of each other. This Hamiltonian could help to describe stack of large aromatic molecules with only one electron in each in the LUMO. It could also be an effective model to describe interacting edge-states in stack of finite graphene planes. This model was already used in \cite{ahn} in the continuum limit, where an interesting new parity effect was pointed out to occur in the dielectric response of one-dimensional stack of quantum rings. Here, we are only interested by the periodicity of the ground state energy with the magnetic flux in the limit of infinite interaction.

Once again, starting from one particular electronic configuration, we are asking the same question about the number of particle hopping in the same direction one should do to come back to it. It is then possible to write down a similar expression than (\ref{periodicity}) with $N_1+N_2+...+N_M=N.M$ and $\alpha=1$ where $M$ is the number of rings and $N$, the number of sites in each ring. Consequently, one can conclude that the periodicity is $1/M$ and that there is no parity effect. The label of the ring acts as a spin degree of freedom. It is worth mentioning that it exists a rigorous mathematical proof of these properties \cite{bernet}.

\subsection{Strongly interacting cylinders}

Last, we are considering a model of interacting cylinders

\begin{equation}
\label{hubbardcr}
\begin{array}{l}
\hat{H}=t\sum_{n,m,\sigma}(a^{\dagger}_{n+1,m,\sigma}a_{n,m,\sigma}e^{i\frac{2\pi}{N}\phi}+h.c.)+t'\sum_{n,m,\sigma}(a^{\dagger}_{n,m+1,\sigma}a_{n,m,\sigma}+h.c.)+\\
\\
\quad \quad \quad \quad U\sum_{n,m,\sigma} \hat{n}_{n,m,\sigma}\hat{n}_{n,m,-\sigma}+U\sum_{n,m,\sigma,\sigma'}\hat{n}_{n,m,\sigma}\hat{n}_{n,m+1,\sigma'}
\end{array}
\end{equation} with again $U \rightarrow \infty$. Again we are asking the same question about the periodicity formulating it in terms of equation (\ref{periodicity}). One has to distinguish between the case of spin $1/2$ and spinless fermions.
\begin{itemize}
\item For spinless fermions, for a non vanishing $t'$, one has $N_1+N_2+...+N_M=N$ and $\alpha=1$; the periodicity is one and there is no parity effect.
\item For spin $1/2$ fermions, whatever is the value of $t'$, one has $N_1+N_2+...+N_M=N.M$ and $\alpha=1$; the periodicity is $1/M$ and there is no parity effect.
\end{itemize}

In conclusion, there is an abrupt transition from a fractional Aharonov-Bohm state to an usual Aharonov-Bohm state going from $t'=0$ to $t' \ne 0$ in the case of spinless fermions; on the contrary, there is no such transition for spin $1/2$ fermions.  

\section{Conclusion}

The main purpose of this review is to point out the strong magnetic field dependence of the static polarisability of nanoscopic cylindrical organic materials when the magnetic field is parallel to the tube axis. This was done by studying two kinds of system.

On one hand we have considered real cylinders in the form of rolled square lattices but also, rolled honeycomb lattices, more specifically, achiral carbon nanotubes. They were studied with both, semiclassical approximation \cite{fulde} and exact diagonalisation \cite{pleutin1}. For all these cases, the polarisability
was shown to present very complex structures as function of the magnetic field
in which one can identify two different
characteristics (cf Fig. 7): (i) the polarisability is a non-continuous function with
sudden jumps separating plateau-like regions (ii) additionally, small peaks
may appear for special values of the magnetic field in place of jumps. A full understanding of these complicated behaviours was given by following
the behaviour of the ground state by increasing the magnetic field:
due to the Aharonov-Bohm effect, many changes of ground state occur and for
some values of the magnetic field accidental degeneracies happen where the
ground state becomes two fold degenerate. A one to one correspondence is found
between the accidents in the polarisability and the accidental degeneracies of
the ground state. Each plateau-like region of the polarisability corresponds
to a quadratic Stark effect with a coefficient proper to the corresponding
magnetic-field induced ground state. Each peak corresponds to a linear Stark
effect appearing at accidental degeneracies when there is direct coupling
between the two states involved (cf Fig. 7).

On the other hand we have studied finite plane of graphene which support edge states. The wave functions of these states have the peculiarity to be mostly localised on the perimeter of the plane defining kind of effective rings. The presence of these states in real materials is a important question since they are able to influence new physical properties \cite{acf} and, we have shown that it is possible to use measurements of the static polarisability in an applied perpendicular magnetic field to detect possible edge states in amorphous carbon materials such as carbon fibres. Moreover we believe that a careful analysis of the quadratic behaviour we have pointed out in section IV (Fig. 11), should give a way to determine the characteristic length of the graphitic region in these disordered systems.

In a last part of this review, we  have briefly described  results of some strongly interacting models in the limit of infinite interactions. The main purpose of this section was to pointed out that the interaction can result in new phenomena such as the fractional Aharonov-Bohm effect which appears for Hubbard rings with spin $1/2$ fermions but also in stacks of uncouple rings with one electron per ring. An explicit treatment of the electron-electron interactions is therefore needed and, this is now our main direction of investigation.

\begin{acknowledgements}
It is a pleasure for us to thank P. Fulde for his constant support, numerous enlightening discussions and careful reading of the manuscript.
\end{acknowledgements}

\begin{figure}
\centerline{\psfig{figure=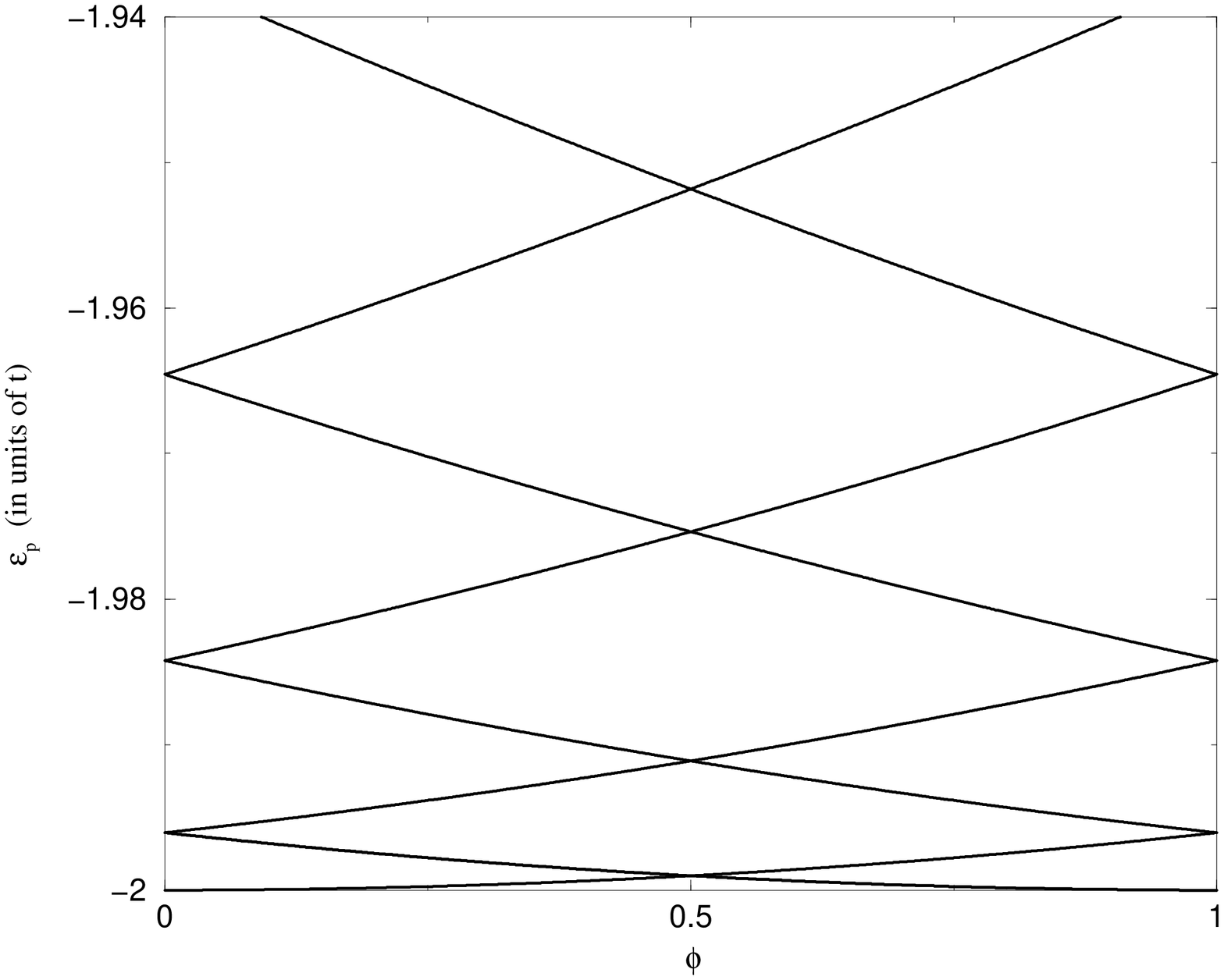,width=15cm,angle=-90}}
\caption{Lowest eigenvalues (eq. (\ref{eigenvaluering})) for a ring with 100 sites as function of flux $\phi$ through the ring (in arbitrary units). Level crossing appear at $\phi=0, 1/2, 1$.}
\label{specring}
\end{figure}

\begin{figure}
\centerline{\psfig{figure=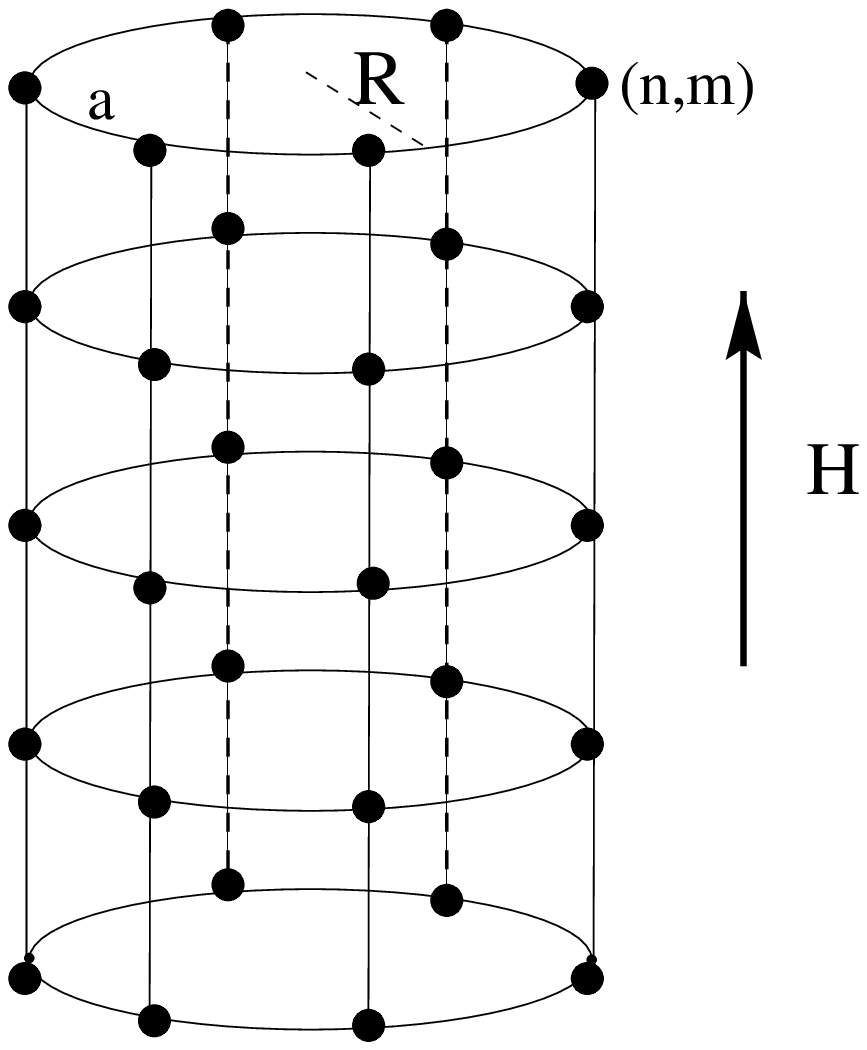,width=10cm}}
\caption{Cylinder made of rolled square lattice placed in a uniform magnetic
fields, $H$, parallel to the cylindrical axis. A uniform electric-field is applied parallel to the cylindrical axis (section III-a) or perpendicular to it (section III-b)}
\label{configuration}
\end{figure}

\begin{figure}
\centerline{\psfig{figure=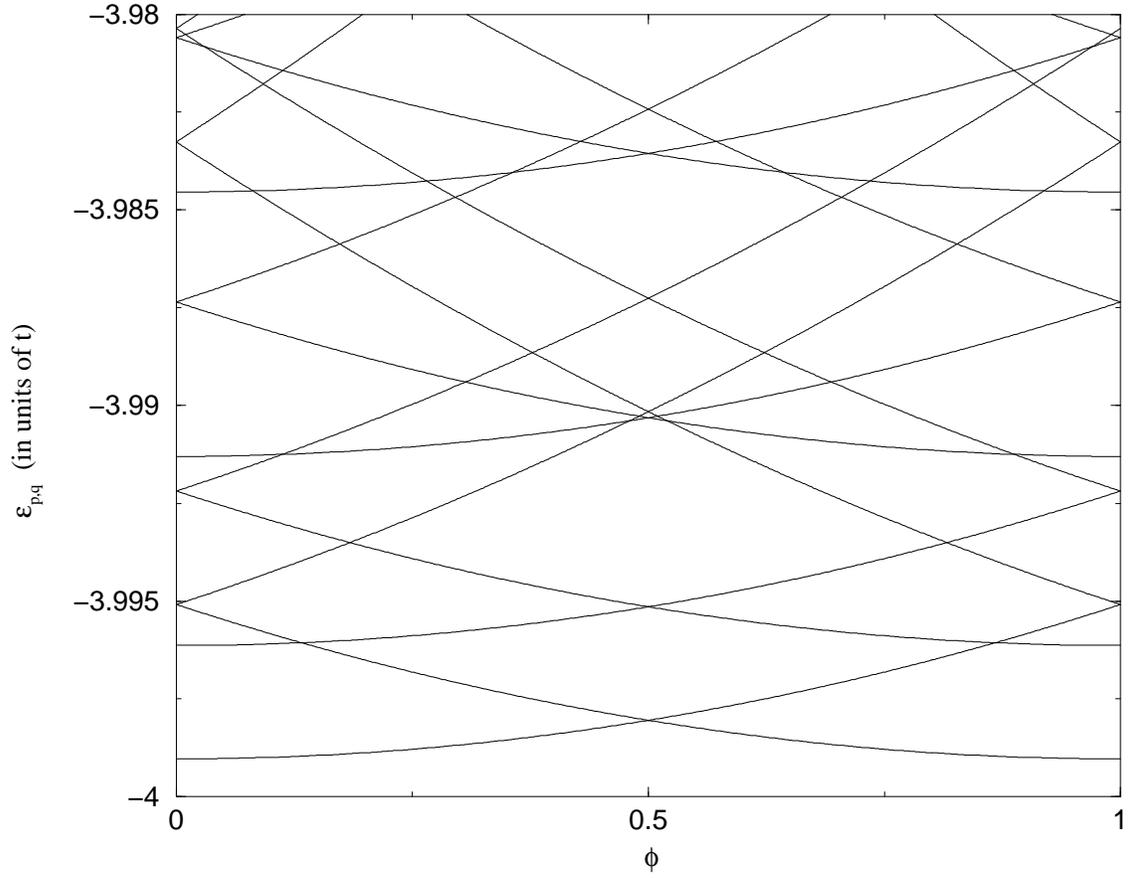,width=15cm,angle=-90}}
\caption{Lowest eigenvalues (Eq. (\ref{spectrum})) for a cylinder with $N=100$ sites along the circumference and $M=100$ sites along the axis, as function of flux $\phi$ through the cylinder (in arbitrary units). Level crossing appear at values of $\phi$ depending on the cylinder geometry.}
\label{speccylinder}
\end{figure}

\begin{figure}
\centerline{\psfig{figure=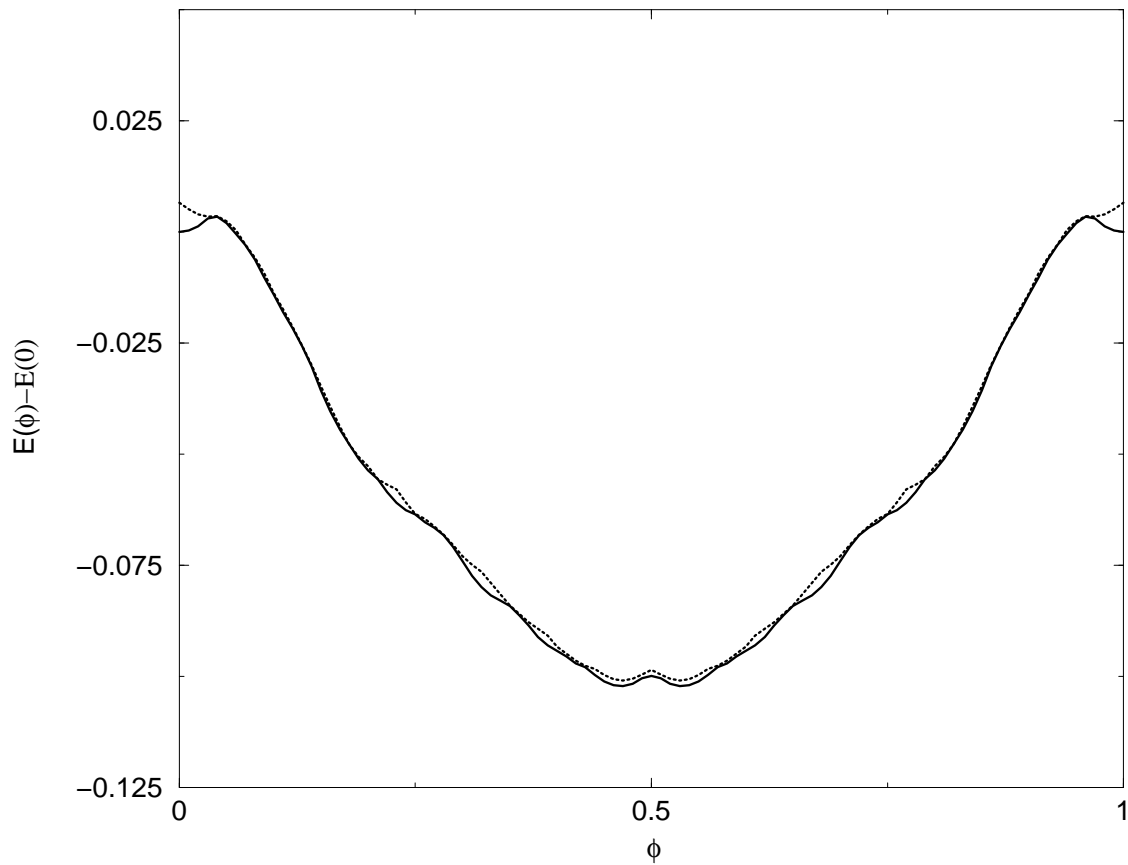,width=15cm,angle=-90}}
\caption{Energies of the ground state (full line) and of the first
excited state (dotted line) as function of the magnetic flux, $\phi$, for a cylinder with $N=100$, $M=100$ and $N_e=1000$ (the references are the energies without magnetic field).}
\label{N100M100Ne1000}
\end{figure}

\begin{figure}
\centerline{\psfig{figure=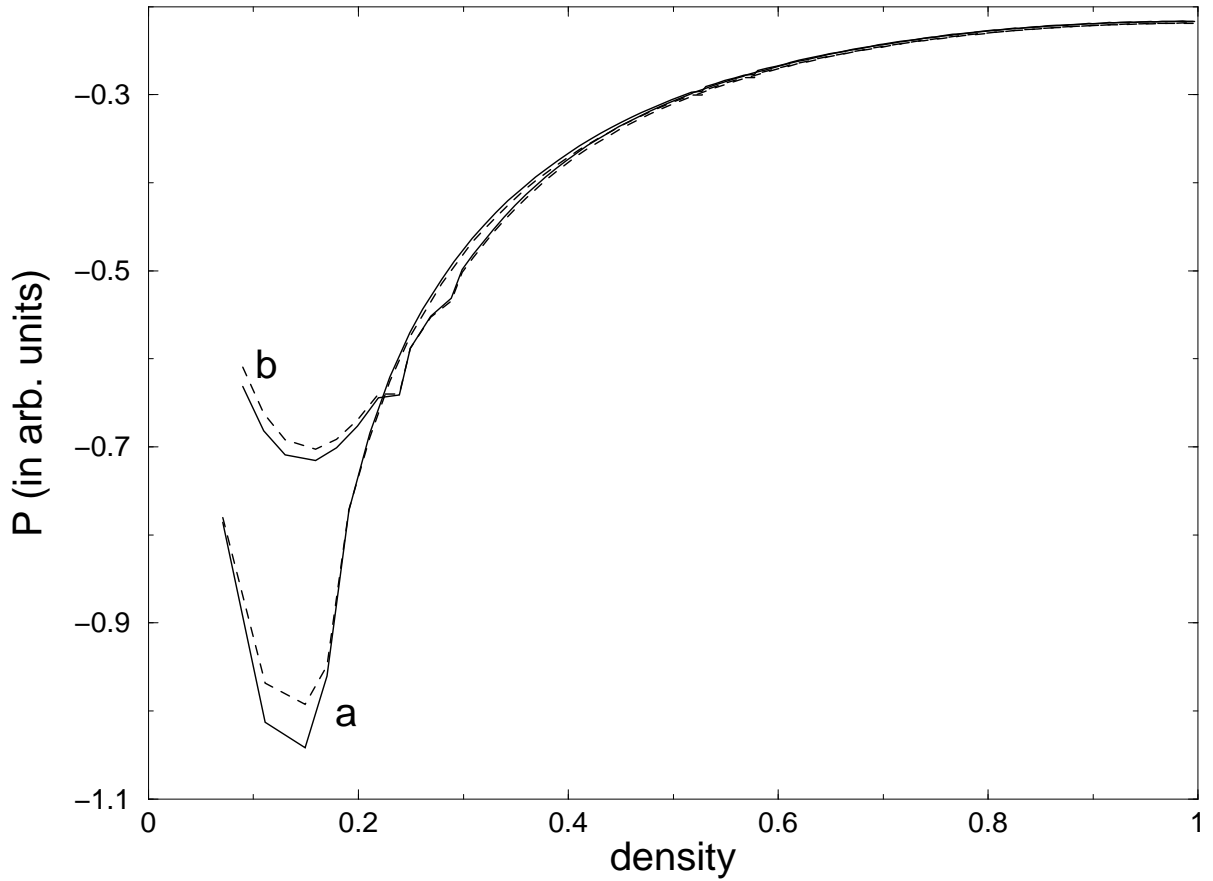,width=15cm,angle=-90}}
\caption{Induced dipole moment as a function of electrons per site for a chain of $N=201$ sites calculated with the exact quantum mechanical expression (\ref{dipole}) (dashed lines) and with the semi-classical expression (\ref{TFChain}) using the Thomas-Fermi approximation (solid lines). (a) and (b) correspond to temperatures $k_BT=0.01$ and $0.05$, respectively (in units of the transfer integral).}
\label{TFring}
\end{figure}

\begin{figure}
\centerline{\psfig{figure=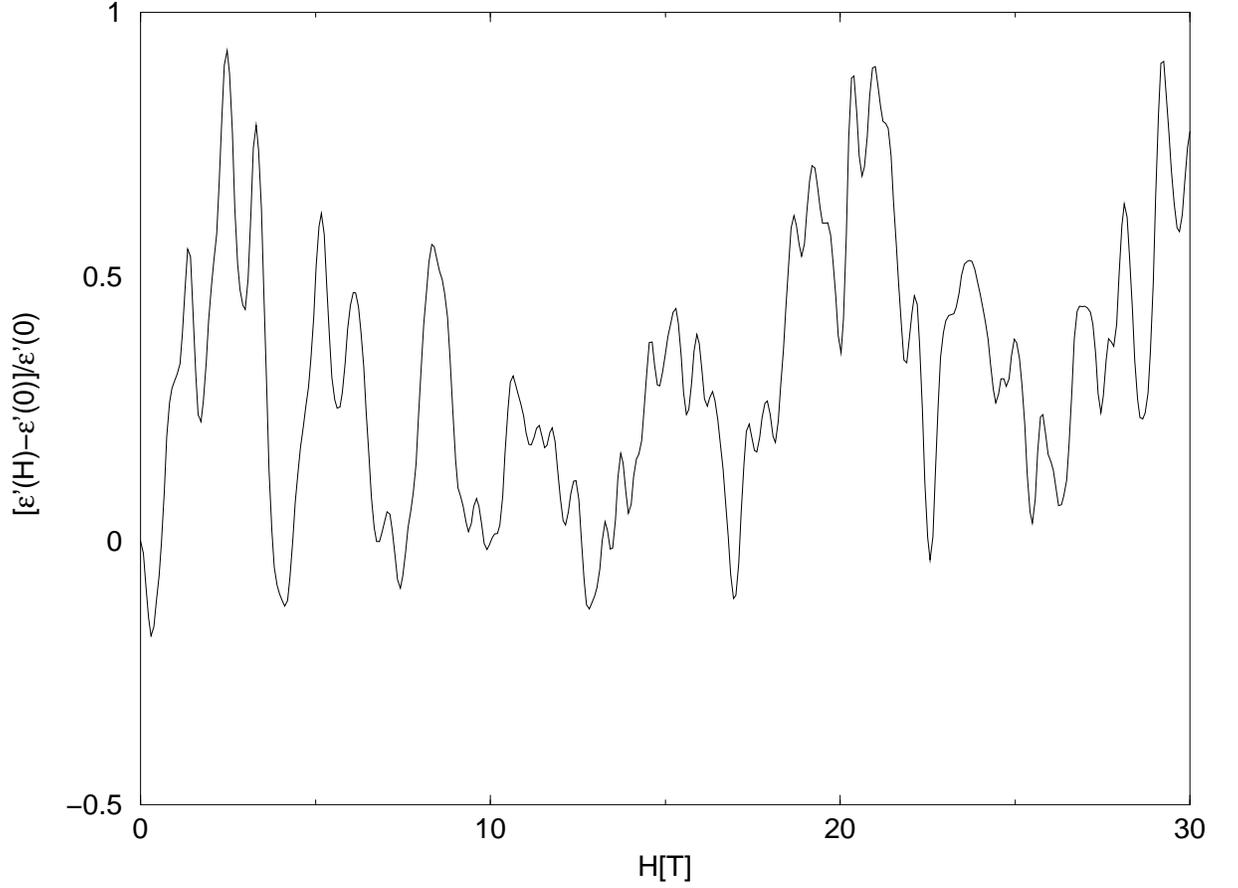,width=15cm,angle=-90}}
\caption{Dielectric response $(\epsilon'(H)-\epsilon'(0))/\epsilon'(0)$ for a model square-lattice system with $N=200$, $M=1000$ in an axial magnetic field. The temperature is $k_BT=10^{-4}$ (in units of the transfer integral), and $a=1.4 \AA$. The electron density is $0.74$ per site.}
\label{TFcylinder}
\end{figure}

\begin{figure}
\centerline{\psfig{figure=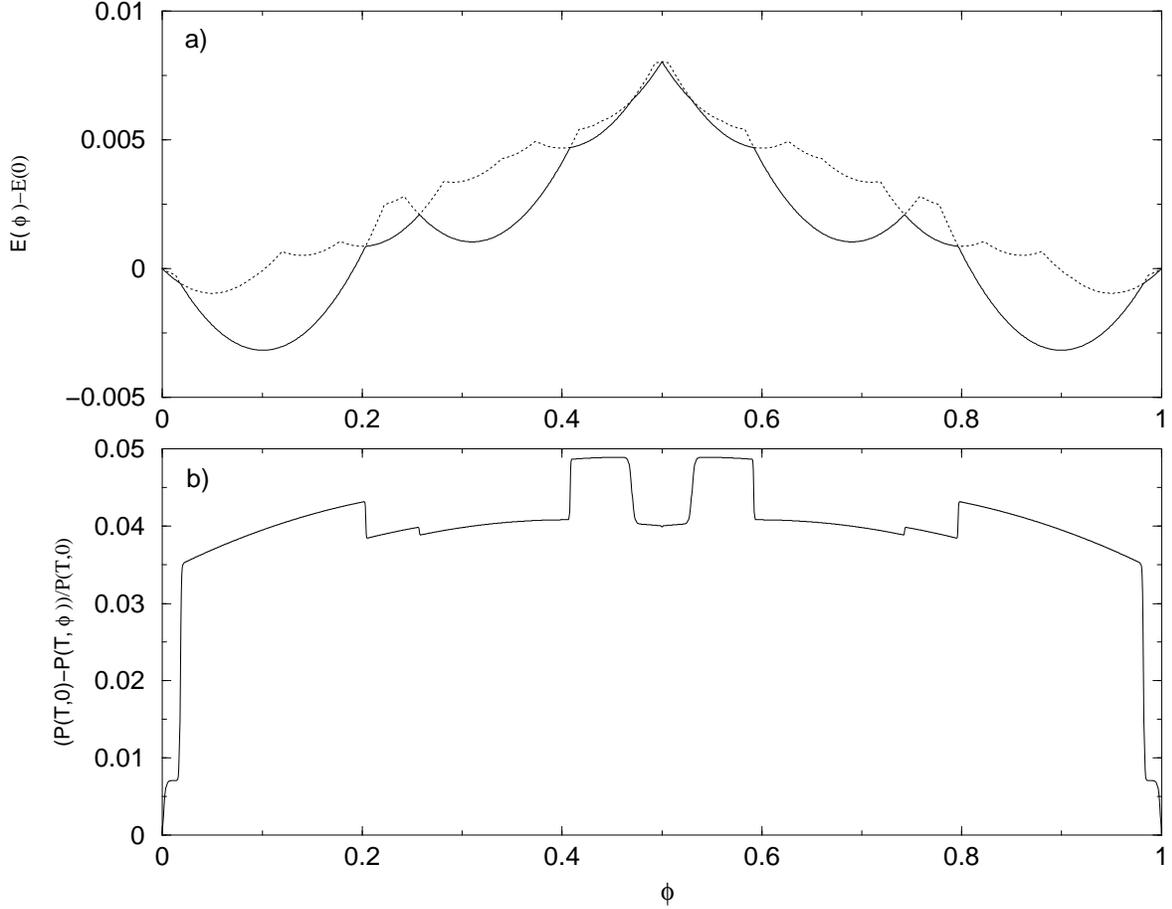,width=15cm,angle=-90}}
\caption{Cylinder with $N=101$, $M=100$ and $N_e=100$, the numbers of site
along the circumference, along the cylindrical axis and number of electrons
(respectively).(a) Energies of the ground state (full line) and of the first
excited state (dotted line) as function of the magnetic flux, $\phi$, without electric
field (the references are the energies without magnetic field). (b) Polarisability $(P(T,0)-P(T,\phi))/P(T,0)$, at $k_BT=10^{-5}t$ and for $v=eER=10^{-3}t$ as
function of the magnetic flux; $t$ is the hopping integral defined in
(\ref{Hcylinder}), $E$ the electric field and $R$ the radius of the cylinder.}
\label{characteristics}
\end{figure}

\begin{figure}
\centerline{\psfig{figure=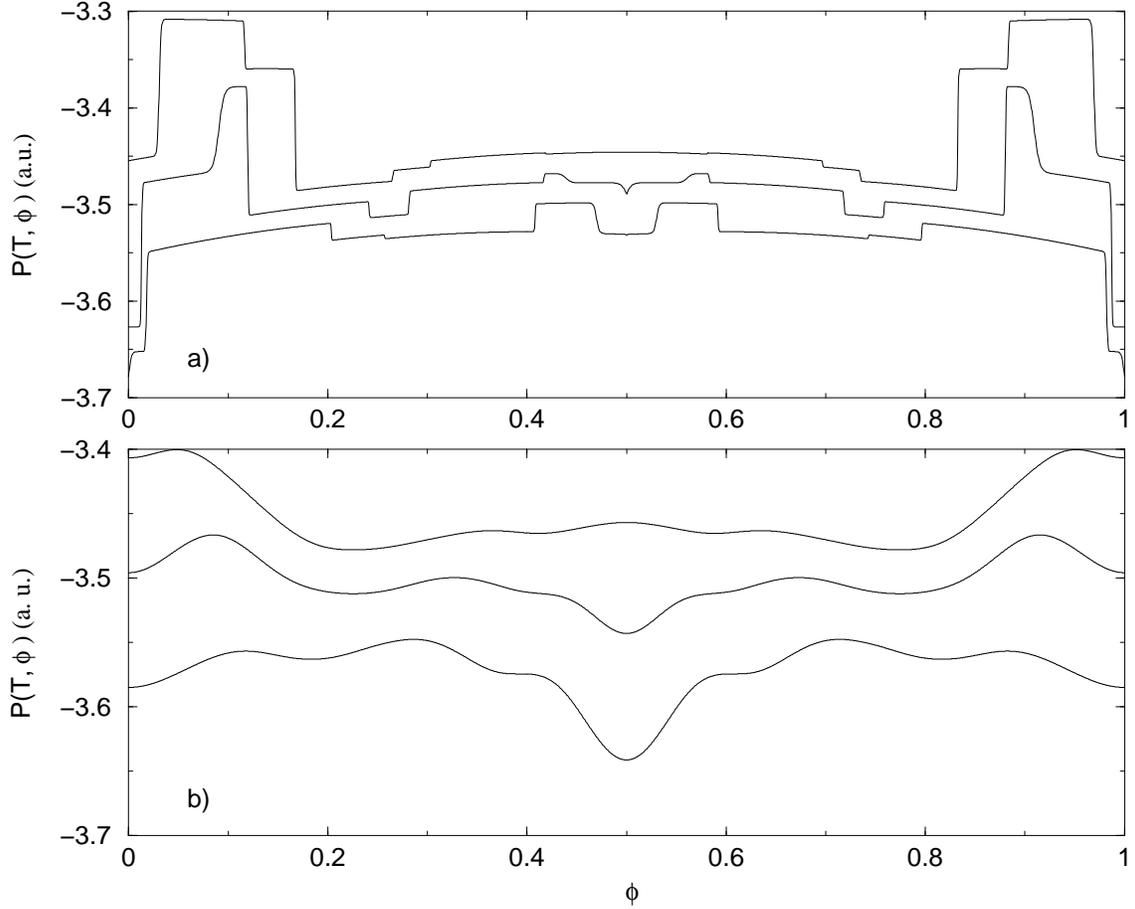,width=15cm,angle=-90}}
\caption{Polarisability of a cylinder with $N=101$ and  $M=100$ as function of
the magnetic flux $\phi$ in arbitrary units (a.u.). (a) For $N_e=100, 101, 102$ from the bottom to the top at a
temperature of $k_BT \simeq 10^{-1}K$. (b) The same but for $k_BT \simeq 10K$.}
\label{fillingeffects}
\end{figure}

\begin{figure}
\centerline{\psfig{figure=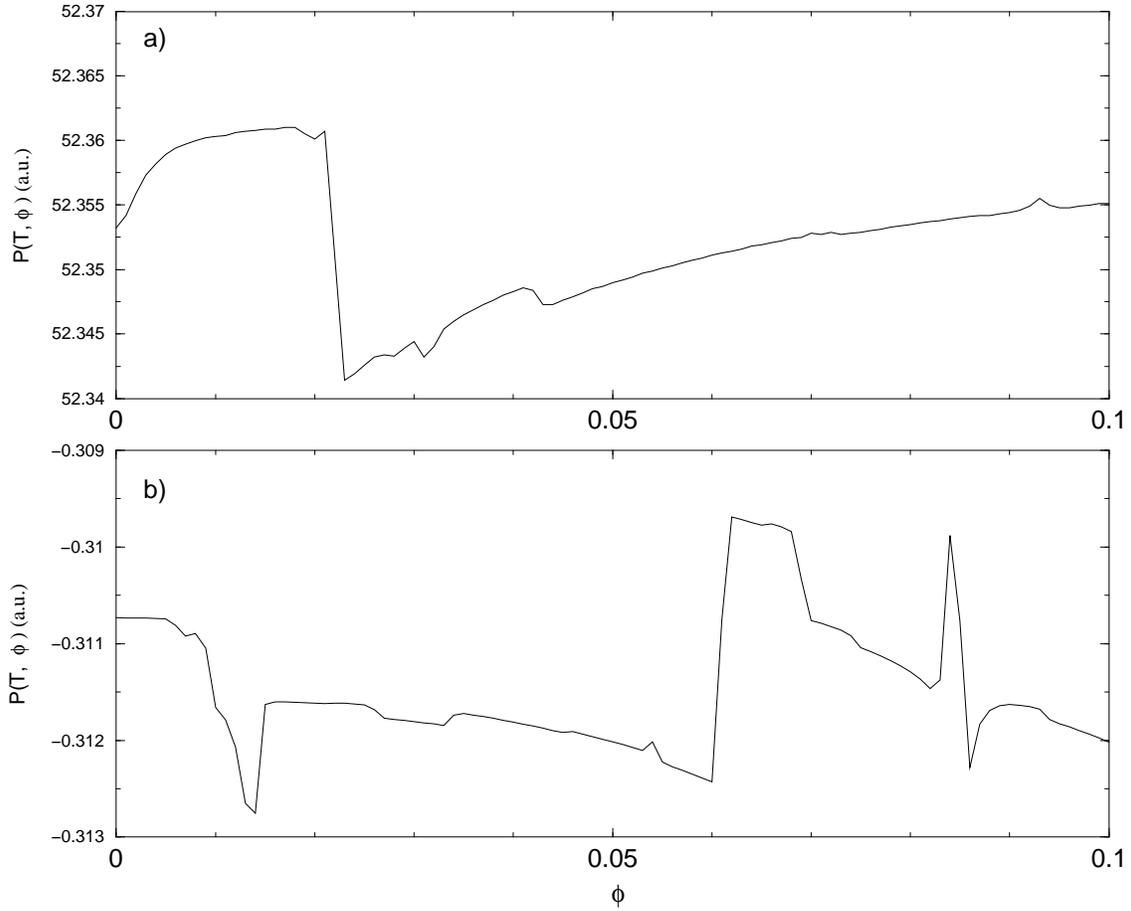,width=15cm,angle=-90}}
\caption{Polarisability in arbitrary units (a.u.) of (a) armchair and (b) zig-zag nanotubes with $N=50$,
$M=500$, $N_e=49000$, $k_BT=10^{-5}t$ and $v=10^{-3}t$ as function of the magnetic flux $\phi$.}
\label{nothalf}
\end{figure}

\begin{figure}
\centerline{\psfig{figure=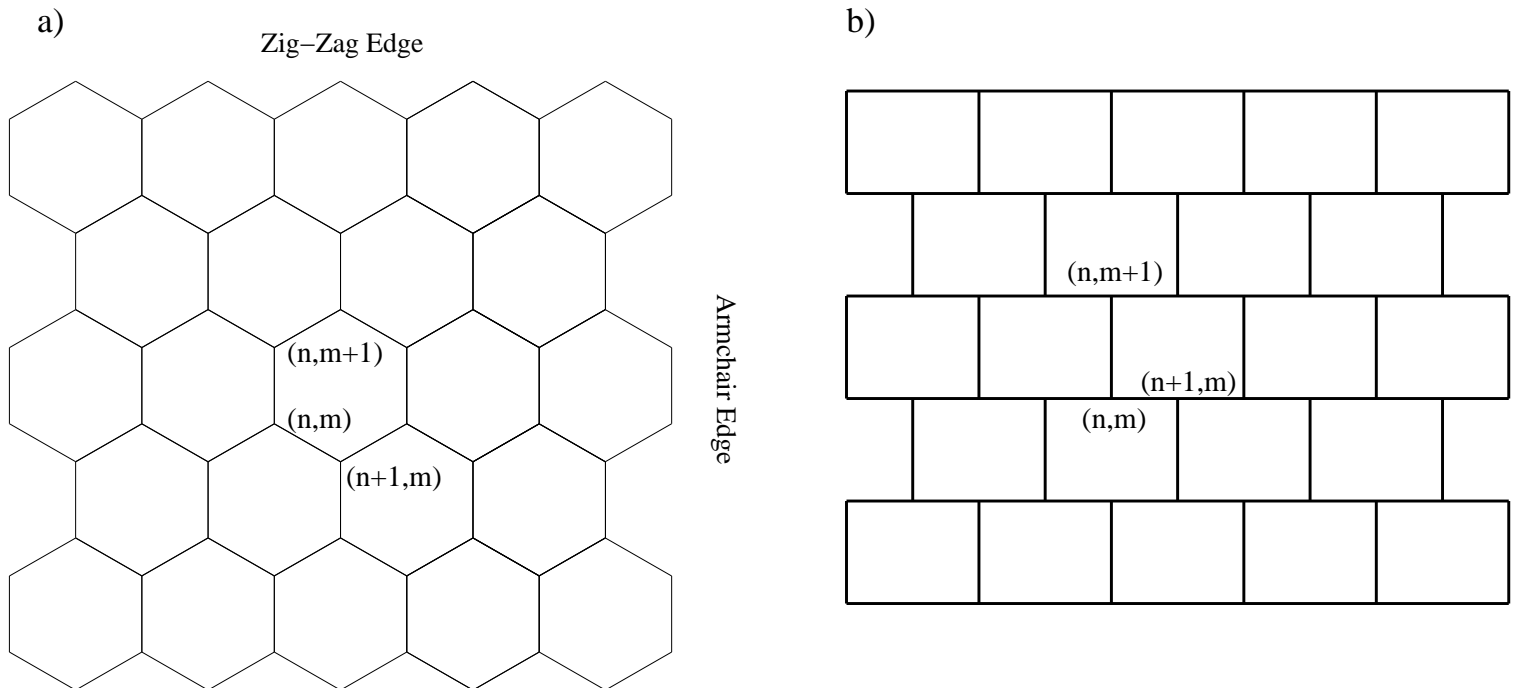,width=20cm}}
\caption{(a) Example of a nanographite plane with rectangular shape. The boundaries have two different kinds of edges: the armchair and the zig-zag edges. The wave function of the edge states are preferentially localised on the zig-zag edges. (b) Corresponding Brick-type lattice equivalent to the honeycomb lattice shown in (a). The sites are labelled with the of (n,m) indices}
\label{honeycomb}
\end{figure}

\begin{figure}
\centerline{\psfig{figure=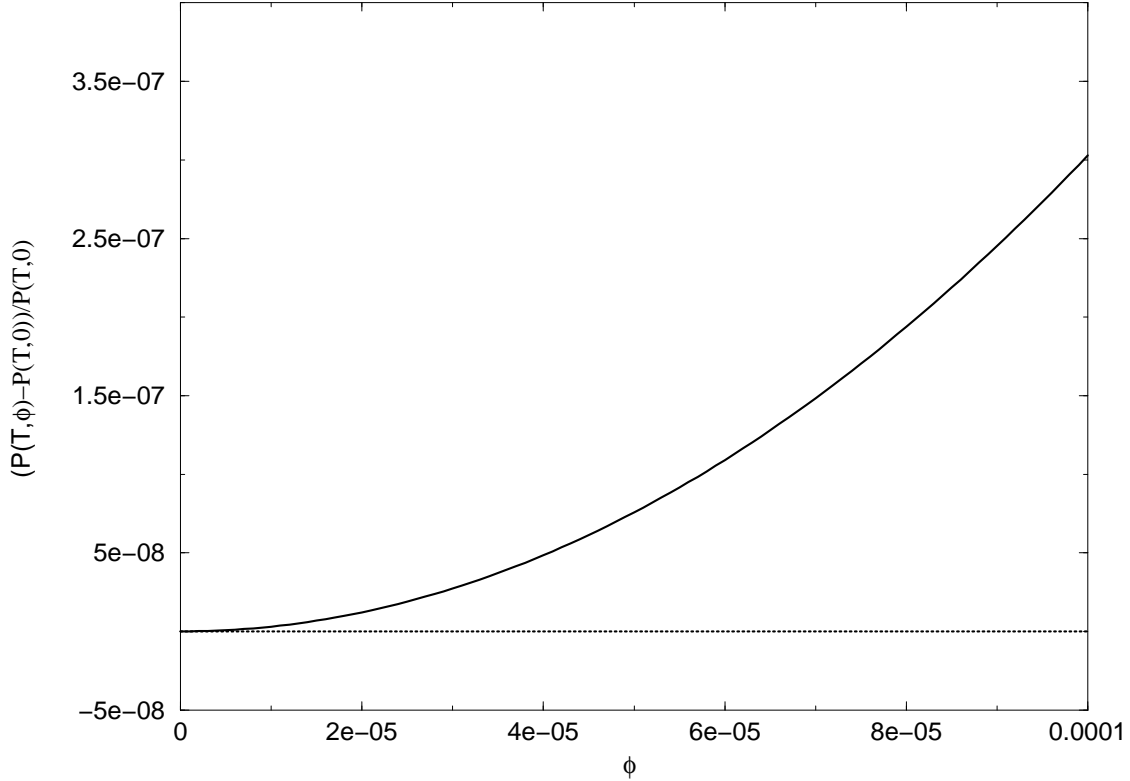,width=15cm,angle=-90}}
\caption{Relative variation of the polarisability, $(P(T,\phi)-P(T,0))/P(T,0)$, of a nanographite plane with 76 benzene rings as a function of the magnetic flux $\phi$, in units of $\phi_0$, for $T\simeq0.1K$. $\phi=10^{-5}$ corresponds to a magnetic field of 1T approximately. The full line is obtained for the case of half-filling; the dotted lines is for the case where we have reduced the electron number by one with respect to half-filling. The edges-states are populated only at half-filling and are responsible for the quadratic dependence seen in this figure.}
\label{poledge}
\end{figure}

\begin{figure}
\centerline{\psfig{figure=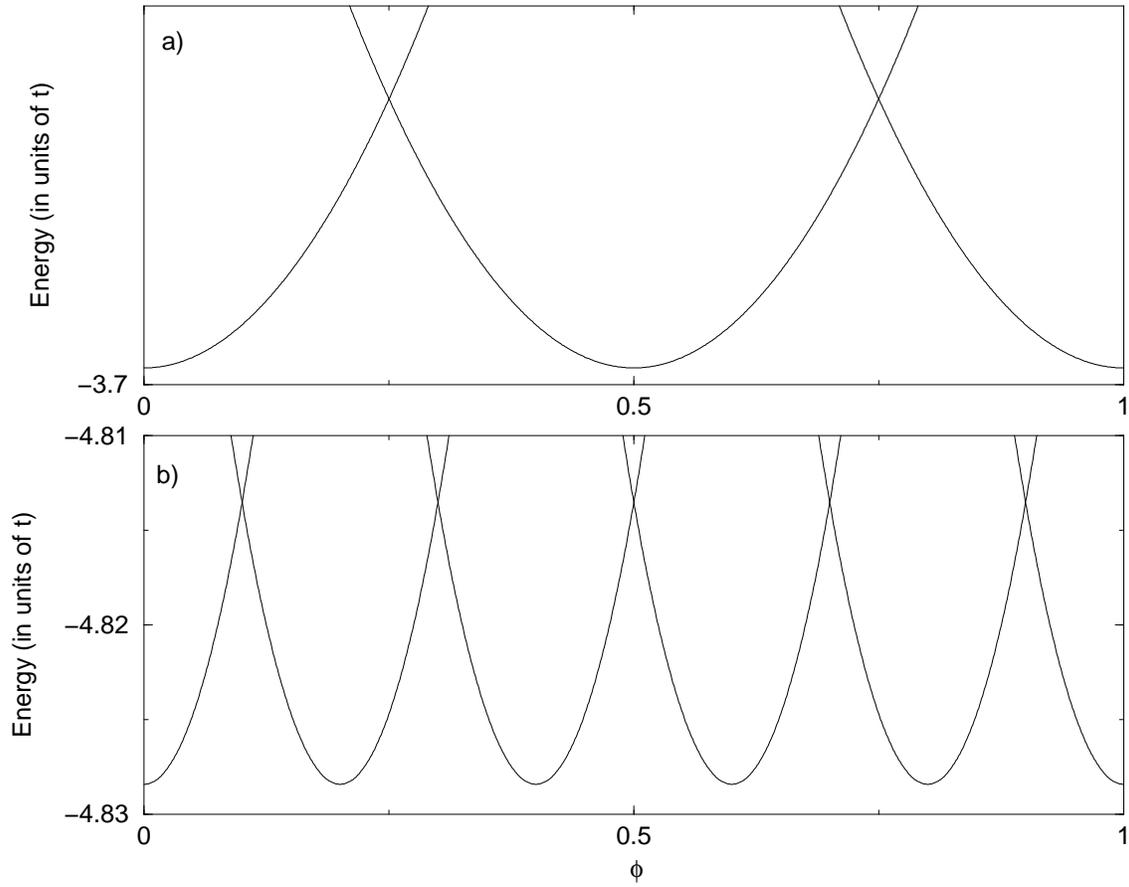,width=15cm,angle=-90}}
\caption{Ground state energy as function of the magnetic flux $\phi$ for Hubbard rings with infinite interaction for a ring with 8 sites and (a) 1 electron up and 1 electron down (b) 3 electrons up and 2 down. In both cases the fractional-Aharonov-Bohm effect is clearly seen.}
\label{FAB}
\end{figure}

\end{document}